\documentclass[preprint]{elsarticle}

\usepackage{graphicx}

\begin{document}

\begin{frontmatter}
\title{Molecular dynamics for long-range interacting systems on Graphic Processing Units}

\author{Tarc\'\i sio M.\ Rocha Filho}
\ead{marciano@fis.unb.br}

\address{Instituto de F\'\i{}sica and International Center for Condensed Matter Physics\\ Universidade de
Bras\'\i{}lia, CP: 04455, 70919-970 - Bras\'\i{}lia, Brazil}

\begin{abstract}
We present implementations of a fourth-order symplectic integrator on graphic processing units for three $N$-body models
with long-range interactions of general interest: the Hamiltonian Mean Field, Ring and two-dimensional
self-gravitating models. We discuss the algorithms, speedups and errors
using one and two GPU units. Speedups can be as high as 140 compared to a serial code, and the overall relative error
in the total energy is of the same order of magnitude as for the CPU code. The number of particles used in the tests
range from 10,000 to 50,000,000 depending on the model.
\end{abstract}

\end{frontmatter}

\section{Introduction}

The dynamics of systems with long-range interactions have been intensively studied over the last two decades due to
their unusual and intriguing phenomenology, such as
the existence of quasi-stationary non-Gaussian states with diverging life-times with the number of particles,
negative microcanonical heat capacity, inequivalence of ensembles and non-ergodicity~\cite{prrev,proc1,proc2,proc3,nos,eplnos,benetti}.
Self-gravitating systems~\cite{1b}, non-neutral plasmas~\cite{1c,1d} and models as the Ring model~\cite{ring1,ring2}
Hamiltonian Mean Field (HMF)~\cite{hmforig},
one-dimensional gravity (infinite uniform density sheets)~\cite{sheet1,sheet2,sheet3}, two-dimensional gravity (infinite uniform rods)~\cite{16b},
Free Electron Laser~\cite{felmod}
and plasma single wave models~\cite{tennyson} are among many examples of systems with long-range forces.
A pair potential interaction has a long-range if it scales for large  distances as $r^{-\alpha }$ with $\alpha <d$, $r$ the
interparticle distance and $d$ the spatial dimension. This slow decaying interparticle potential is responsible for the coupling of distant
components of the system in such a way that all particles contribute to the dynamics of a given particle.
For out of equilibrium situations, many of these studies rely on molecular dynamics simulations, i.~e.\ solving numerically the Hamiltonian
equations of motion for the $N$-particle system. It is also a well known fact that, under suitable conditions, the statistical description of
the dynamics of long range interacting systems is equivalent to the Vlasov equation in the $N\rightarrow\infty$ limit~\cite{prrev,braun}.

Here we present a CUDA implementation~\cite{cuda} of Molecular Dynamics (MD) algorithms on Graphics Processing Units (GPUs) to solve
the Hamiltonian equations of motion for such systems, using one and two GPUS. The algorithms used
can be efficiently extended for any number of GPUs. Molecular Dynamics simulations have been extensively used
to study many properties of these systems from first principles (see for instance Reference~\cite{prrev} and references therein).
Some numerical parallel algorithms were implemented in the literature with applications ranging from condensed matter
to astrophysics~\cite{moore,trott,bedorf,rapaport,morozov,sunarso}. Although MD simulations are widely used in the study of long-range systems,
only recently a GPU code of a symplectic integrator was implemented and used by the author and collaborators alongside a GPU Vlasov
solver to study non-equilibrium phase transitions in the HMF model~\cite{nos2,nos3,eu1}. In such studies simulations are performed without introducing
any simplifying hypothesis such as the Particle-Mesh technique~\cite{fellhauer} or similar methods commonly used in condensed matter physics~\cite{baker},
with a full N-body method with a computational effort scaling as $N^2$. For some models the explicit form of the pair interaction potential allows
further simplifications with important reduction in computational time.

The structure of the paper is the following: in section~\ref{mlri} we describe the model systems that are studied in this paper.
Section~\ref{algcuda} presents the main algorithms and how they are implemented in CUDA, and in section~\ref{resdis}
we present the timings and speedups for the three models taken as examples, and discuss how the relative error in the energy
behave for each of them. We close the paper with some concluding remarks in section~\ref{conc}.

\section{Models with long-range interactions}
\label{mlri}

Here we present some simplified models that are discussed in the present work.
The Ring model is composed of $N$ particles with unit mass on a ring of radius $R$ and interacting though a regularized gravitational
potential with Hamiltonian~\cite{ring1,ring2}:
\begin{equation}
H=\frac{1}{2}\sum_{i=1}^N p_i^2-\frac{1}{N}\sum_{i<j=1}^N \frac{1}{\sqrt{2}\sqrt{1-\cos(\theta_i-\theta_j)+\epsilon}},
\label{ringham}
\end{equation}
with $\epsilon$ a softening parameter introduced to avoid the divergence at zero distance
and $\theta_i$ denotes the position angle of a particle on the circle.
By considering the limit for large values of the parameter $\epsilon$ we obtain the Hamiltonian Mean Field (HMF) model with
Hamiltonian~\cite{hmforig}:
\begin{equation}
H=\frac{1}{2}\sum_{i=1}^N p_i^2+\frac{1}{N}\sum_{i<j=1}^N\left[1-\cos\left(\theta_i-\theta_j \right)\right].
\label{hmfham}
\end{equation}

The two-dimensional self-gravitating particles is composed of identical particles
with unit mass. By solving the Poisson equation in
two dimensions we obtain the Hamiltonian~\cite{16b}:
\begin{equation}
H=\frac{1}{2}\sum_{i=1}^N\:(p_{x,i}^2+p_{y,i}^2)+\frac{1}{N}\sum_{i<j=1}^N\:\log\left[(x_i-x_j)^2+(y_i-y_j)^2+\epsilon\right],
\label{2dham}
\end{equation}
where $p_{x,i}$, $p_{y,i}$, $x_i$ and $y_i$ are the $x$ and $y$ components of the momentum and the coordinates for the $i$-th particle,
respectively, and $\epsilon$ is again a (small) parameter used to avoid the divergence of the potential at zero distance.

\section{Algorithms and CUDA implementation}
\label{algcuda}

The standard integration method for the numerical solution of Hamilton equations for the model systems described in the previous section
is a fourth-order symplectic integrator. Here we adopt the Yoshida symplectic integrator that approximates the evolution operator as~\cite{yoshida}:
\begin{eqnarray}
\lefteqn{e^{\Delta t\hat L_H}=e^{B_0\Delta t\hat L_V}e^{D_0\Delta t\hat L_K}e^{B_1\Delta t\hat L_V}}\nonumber\\
 & & \times e^{D_1\Delta t\hat L_K}e^{B_1\Delta t\hat L_V}
e^{D_0\Delta t\hat L_K}e^{B_0\Delta t\hat L_V}+{\cal O}\left(\Delta t^5\right),
\label{yoshint}
\end{eqnarray}
where $\Delta t$ is the integration step and $B_0$, $B_1$, $D_0$ and $D_1$ are numerical constants obtained in Ref.~\cite{yoshida},
$\hat L_H\equiv\{H,\:\:\:\}$, $\hat L_V\equiv\{V,\:\:\:\}$ and $\hat L_K\equiv\{K,\:\:\:\}$ with $K$ and $V$ the kinetic and potential energies
respectively and $\{H,\:\:\:\}$ stands for the Poisson bracket of $H$ with any function it operates on. The approximation in eq.~(\ref{yoshint})
is time reversible. A whole integration step is given by the following steps where ${\bf f}({\bf r})$ stand for the force array at positions
${\bf r}_1,\ldots,{\bf r}_N$:
\begin{enumerate}
\item ${\bf f}^{(I)}={\bf f}({\bf r}(t))$,
\item ${\bf p}^{(I)}={\bf p}(t)+B_0\Delta t\:{\bf f}^{(I)}$,
\item ${\bf r}^{(I)}={\bf r}(t)+D_0\Delta t\:{\bf p}^{(I)}$,
\item ${\bf f}^{(II)}={\bf f}({\bf r}^{(I)})$,
\item ${\bf p}^{(II)}={\bf p}^{(I)}+B_1\Delta t\:{\bf f}^{(II)}$,
\item ${\bf r}^{(II)}={\bf r}^{(I)}+D_1\Delta t\:{\bf p}^{(II)}$,
\item ${\bf f}^{(III)}={\bf f}({\bf r}^{(II)})$,
\item ${\bf p}^{(III)}={\bf p}^{(II)}+B_1\Delta t\:{\bf f}^{(III)}$,
\item ${\bf r}(t+\Delta t)={\bf r}^{(II)}+D_0\Delta t\:{\bf p}^{(III)}$,
\item ${\bf f}(t+\Delta t)={\bf f}({\bf r}(t+\Delta t))$,
\item ${\bf p}(t+\Delta t)={\bf p}^{(III)}+B_0\Delta t\:{\bf f}^{(I)}$,
\item Compute any properties of interest, then go to step (2) with ${\bf f}^{(II)}={\bf f}(t+\Delta t)$.
\end{enumerate}
Each time step requires three force calculations, being the most demanding step.
We now present the specific CUDA implementation for each model in section~\ref{mlri}.

\subsection{HMF model}

From the Hamiltonian in eq.~(\ref{hmfham}) the potential energy and the force on particle $i$ can be written respectively as
\begin{equation}
V=\frac{N}{2}\left[1-M_x^2-M_y^2\right],
\label{hmfpot}
\end{equation}
and
\begin{equation}
f_i=\cos(\theta_i)M_y-\sin(\theta_i)M_x,
\label{hmfforce}
\end{equation}
where the components of the ``magnetization'' are given by
\begin{equation}
M_x=\frac{1}{N}\sum_{i=1}^N\cos(\theta_i),\hspace{5mm}M_y=\frac{1}{N}\sum_{i=1}^N\sin(\theta_i).
\label{magnetdef}
\end{equation}
We note that due to the form of the interaction potential the symplectic integration time for the HMF scales with $N$ in contrast to the
usual $N^2$ scaling for the other two models below. The most demanding part of an integration step is the computation of the sine and cosine
of the position angles for each particle. For the force in eq.~(\ref{hmfforce}) each $\cos(\theta_i)$ and $\sin(\theta_i)$ is used twice:
to compute $M_x$ and $M_y$ and to obtain the force on each particle from expression~(\ref{hmfforce}).
To avoid redundant computations their values are first computed and stored in an array
in the GPU global memory taking care to ensure coalesced memory access, which is an important issue in any CUDA implementation.
A CUDA kernel is composed of blocks each with a given number of threads. Each thread in a block computes the value of the cosine and sine of
a given particle and the corresponding values are also stored in shared memory. A reduction procedure is then used to compute the values of
$M_x$ and $M_y$ and the force is obtained using the precomputed values of $\sin(\theta_i)$ and $\cos(\theta_i)$.
The potential energy is trivially obtained from eq.~(\ref{hmfpot}) and the kinetic energy is efficiently computed using
a straightforward reduction procedure.

For two GPUs half of particle data (position, momenta and force) is stored in each GPU which then computes the magnetization components for its
corresponding number of particles, and their sum give the total magnetization components. The forces on each particle are the
trivially computed for the particles on each GPU, and the subsequent evolution is also performed independently by each GPU for its set of particles.

\subsection{Ring model and self-gravitating 2D system}

\begin{figure}[ptb]
\begin{center}
\scalebox{0.7}{{\includegraphics{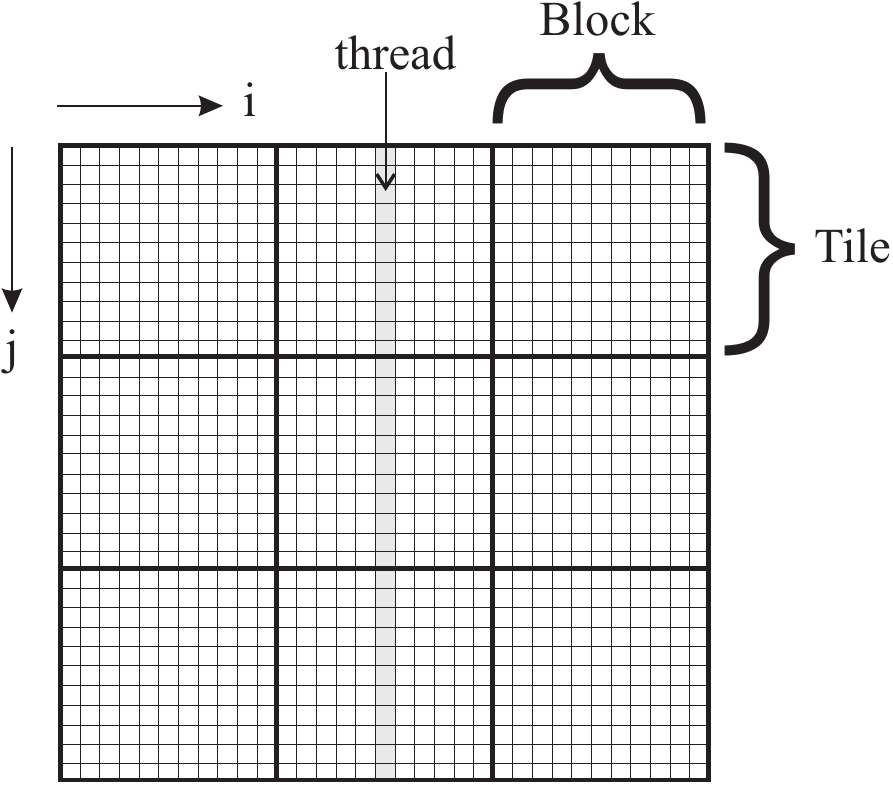}}}
\end{center}
\caption{Decomposition in tiles for the computation of the force components $f_i$ for the Ring and self-gravitating models.}
\label{fig1}
\end{figure}

The computational time for symplectic integration of the Ring model scales as $N^2$.
Here we follow a strategy similar to the one described in Ref.~\cite{nyland} based on a decomposition of the force calculation in tiles
as depicted in Figure~\ref{fig1}. The force on particle $i$ is obtained from Hamiltonian~(\ref{ringham}) as:
\begin{equation}
f_i=\sum_j f_{ij},\hspace{5mm}f_{ij}=\frac{1}{2\sqrt{2}N}\frac{\sin(\theta_i-\theta_j)}{\left(1-\cos(\theta_i-\theta_j)+\epsilon\right)^{3/2}}.
\label{ringforce}
\end{equation}
This last expression can be rewritten as
\begin{equation}
f_{ij}=\frac{1}{2\sqrt{2}N}\frac{\sin\theta_i\cos\theta_j-\cos\theta_i\sin\theta_j}
{\left(1-\cos\theta_i\cos\theta_j-\sin\theta_i\sin\theta_j+\epsilon\right)^{3/2}}.
\label{fijsimp}
\end{equation}
Each $\sin\theta_i$ and $\cos\theta_i$ for $i=0,\ldots,N$ is computed and stored in global memory in order to avoid computing twice their values.
Each tile has $N_{\rm threads}$ (the number of threads in a block)
particles in the horizontal direction (index $i$ in Fig.~\ref{fig1}) and $N_{\rm threads}$ in the vertical direction (index $j$).
The total number of tiles in each direction is thus $N_{tiles}=N/N_{\rm threads}$. The algorithm can be expressed as:
\begin{enumerate}
\item Store $f_k=0$ and the values of $\sin\theta_k$ and $\cos\theta_k$,
$k=l\:N_{\rm threads},\ldots,(l+1)\:N_{\rm threads}-1$ in shared memory, with $l$ the block number,
and synchronize threads in the block.
\item $m=0$;
\item Store $\sin\theta_k$ and $\cos\theta_k$, $k=m\:N_{\rm threads},\ldots,(m+1)\:N_{\rm threads}-1$ in shared memory and synchronize threads in the block;
\item for $k=m\:N_{\rm threads},\ldots,(m+1)\:N_{\rm threads}-1$ compute $f_{ik}$ and sum its values to $f_i$;
\item $j=j+1$ and goto step (3) while $j<N_{tiles}$.
\end{enumerate}
Newton's third law is not used here as its implementation in a CUDA kernel would introduce unnecessary complications with no significant speed gain.
The computation of $1/\sqrt{x}$ is performed using the CUDA function rsqrt(x) with significant gain in computing time
and no significant loss in accuracy.

For the two-dimensional self-gravitating system the approach is essentially the same as for the Ring model described above,except for the number of components
for each particle (two) and the interpaticle force. In this case, a significant gain in speed without significantly compromising accuracy consists in
using double precision for the storage of all data in global memory but storing each component of the position in single precision shared memory
before computing the force on each particle,
which allows to load more blocks concomitantly on the GPU, and therefore augmenting occupancy.
The force between two given particles is then computed in single precision but added in a double precision variable to determine the total force on a given particle.
All remaining computations are performed in double precision. For determining the timings and for comparison purposes the same is done for the CPU code.

\begin{figure}[ptb]
\begin{center}
\scalebox{0.7}{{\includegraphics{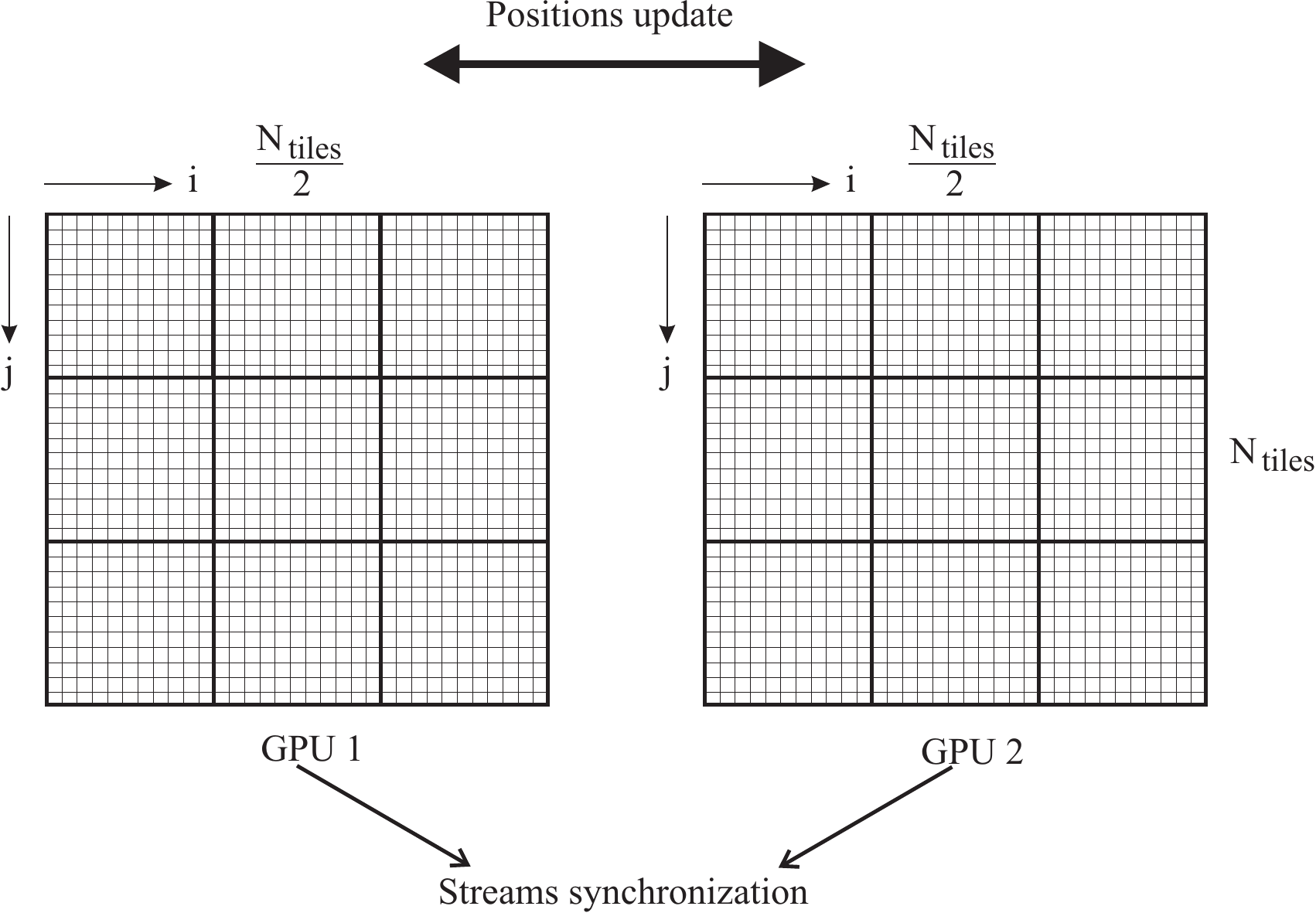}}}
\end{center}
\caption{Using two GPUs to compute the forces for the Ring model and 2D self-gravitating system.}
\label{fig2}
\end{figure}

For two GPUs, each one computes separately the forces for half of the particles, which contrary to the HMF model requires the positions for all particles.
Therefore each GPU stores in global memory half on the momenta, half of the forces but all the particle position coordinates. As for the HMF model
the time evolution is performed independently on each GPU for its respective set of particles, and only half of the positions on each GPU is up to date after
a time step. To compute the force we first copy half of the updated values from one GPU to the other using an asynchronous memory copy.
Then each GPU computes half of the forces with $N_{\rm tiles}/2$ tile in the horizontal and $N_{\rm tiles}$ tiles in the vertical directions.
The whole computation is synchronized after all forces are evaluated
using stream synchronization, as shown in Fig.~\ref{fig2}. A single partial integration step composed of a force
calculation and free drift with constant momentum and a force increment with constant position
(steps 1, 2 and 3 in section~\ref{algcuda} for instance) can be summarized for both the Ring and
2D self-gravitating model as (each GPU has a stream defined for it):
\begin{enumerate}
\item Asynchronous copy of the position components of particles $1,\ldots,N/2$ from GPU 0 to GPU 1.
\item Asynchronous copy of the position components of particles $N/2+1,\ldots,N$ from GPU 1 to GPU 0.
\item Stream synchronization for each GPU.
\item Launch a kernel to compute the components of ${\bf f}_i$, $i=1,\ldots,N/2$.
\item Launch a kernel to compute the components of ${\bf f}_i$, $i=N/2+1,\ldots,N$.
\item Stream synchronization for each GPU.
\item Launch a kernel to update halt of the momenta and half of the position coordinates on GPU 0.
\item Launch a kernel to update halt of the momenta and half of the position coordinates on GPU 1.
\item Stream synchronization for each GPU.
\end{enumerate}
The asynchronous copy between GPUs in steps (1) and (2) above is efficiently handled by the CUDA routine {\tt cudaMemcpyPeerAsync}.

\section{Results and discussion}
\label{resdis}

The computer used for the simulations is an i7-2600/ 3.40 GHz and 16 GB of RAM.
The GPU is a GeForce GTX 690 dual with a Kepler Architecture, 2048 MBytes of memory and
1536 CUDA Cores for each GPU. Results are presented using one and both units Tables~\ref{tab1}, \ref{tab2} and~\ref{tab3}
show the timings for a complete time step for the HMF, Ring and 2D gravity models, respectively. All CPU implementations were optimized
on a single CPU core but parallel implementations on many cores on the CPU are also possible with a maximum speedup given by the number of cores.
The speedups obtained range from 23 to 73 for the HMF model, 104 to 141 for the Ring model and 53 to 100 for the 2D self-gravitating system.
The greater the number of particles the greater the speedup due to a higher occupancy of the GPU cores. For the same reason
using more than one GPU becomes more advantageous for a large number of particles.  It is interesting to note that the smaller speedups were
obtained for the HMF model. A similar result is obtained in Ref.~\cite{eu1} for the solution of the Vlasov equation
and is a consequence of the scaling property with $N$ of the computational time
and the fact the CPU code being already highly optimized.

\begin{figure}
\begin{center}
\scalebox{0.25}{{\includegraphics{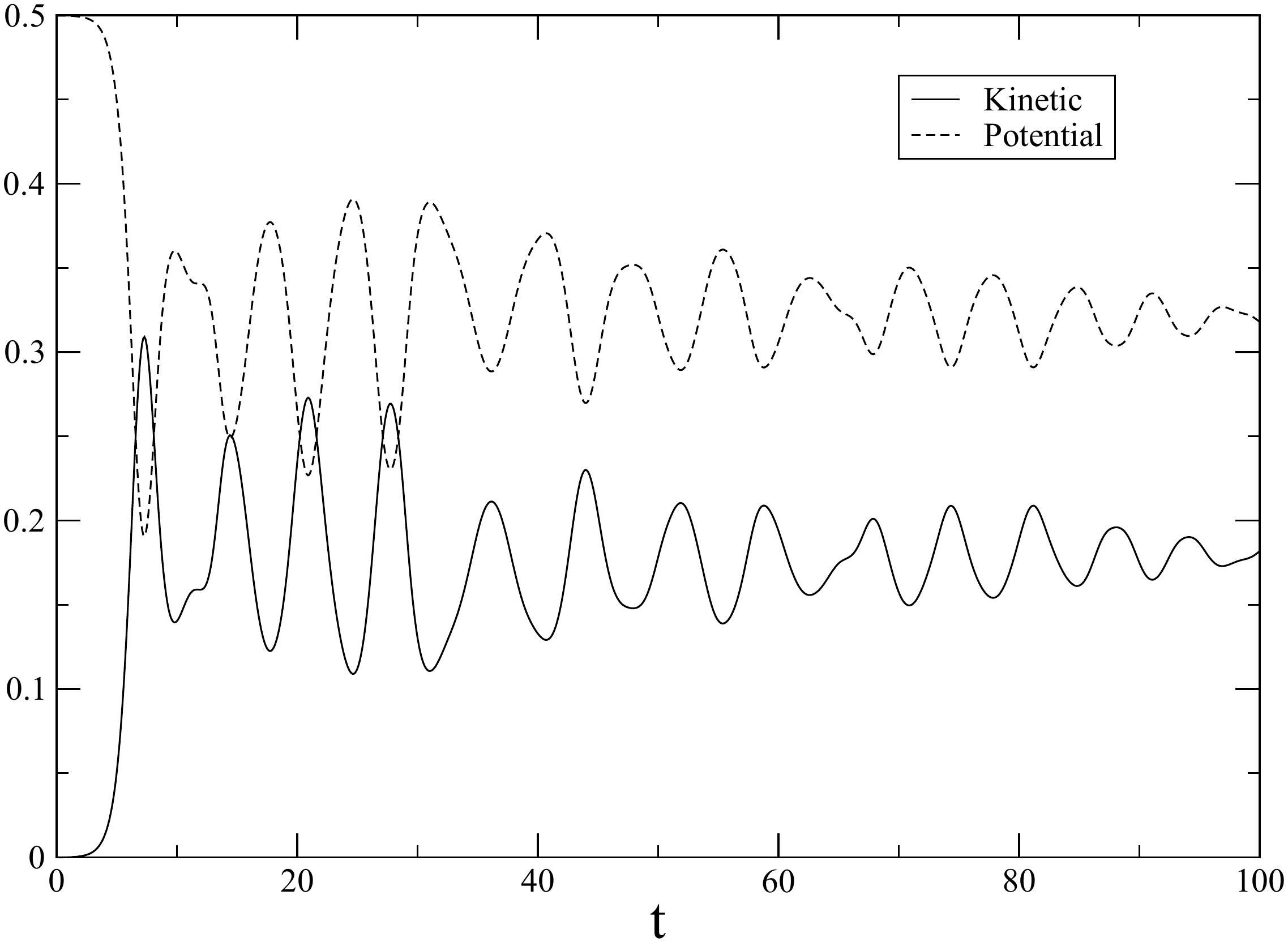}}}
\scalebox{0.25}{{\includegraphics{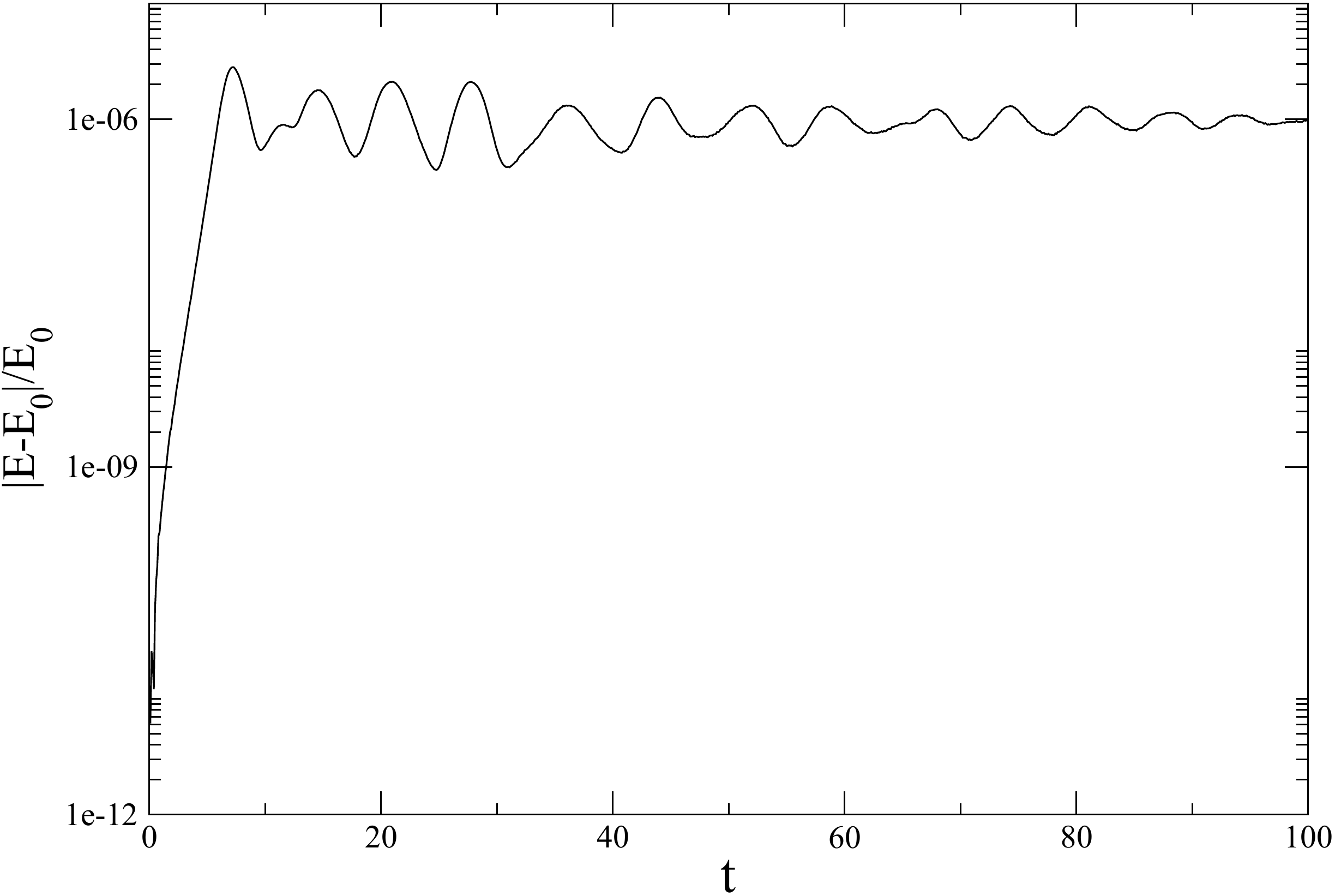}}}
\end{center}
\caption{Left panel: Kinetic and potential energy as a function of time for the HMF model. Right panel:
Relative error for the CUDA implementation for the HMF model. The error for the CPU versions is virtually the same and would
be indistinguishable on the graphic. The simulation data are $\epsilon=10^{-3}$ and $\Delta t=0.1$ for a homogeneous initial condition with particles at rest.}
\label{fig3}
\end{figure}

\begin{figure}
\begin{center}
\scalebox{0.25}{{\includegraphics{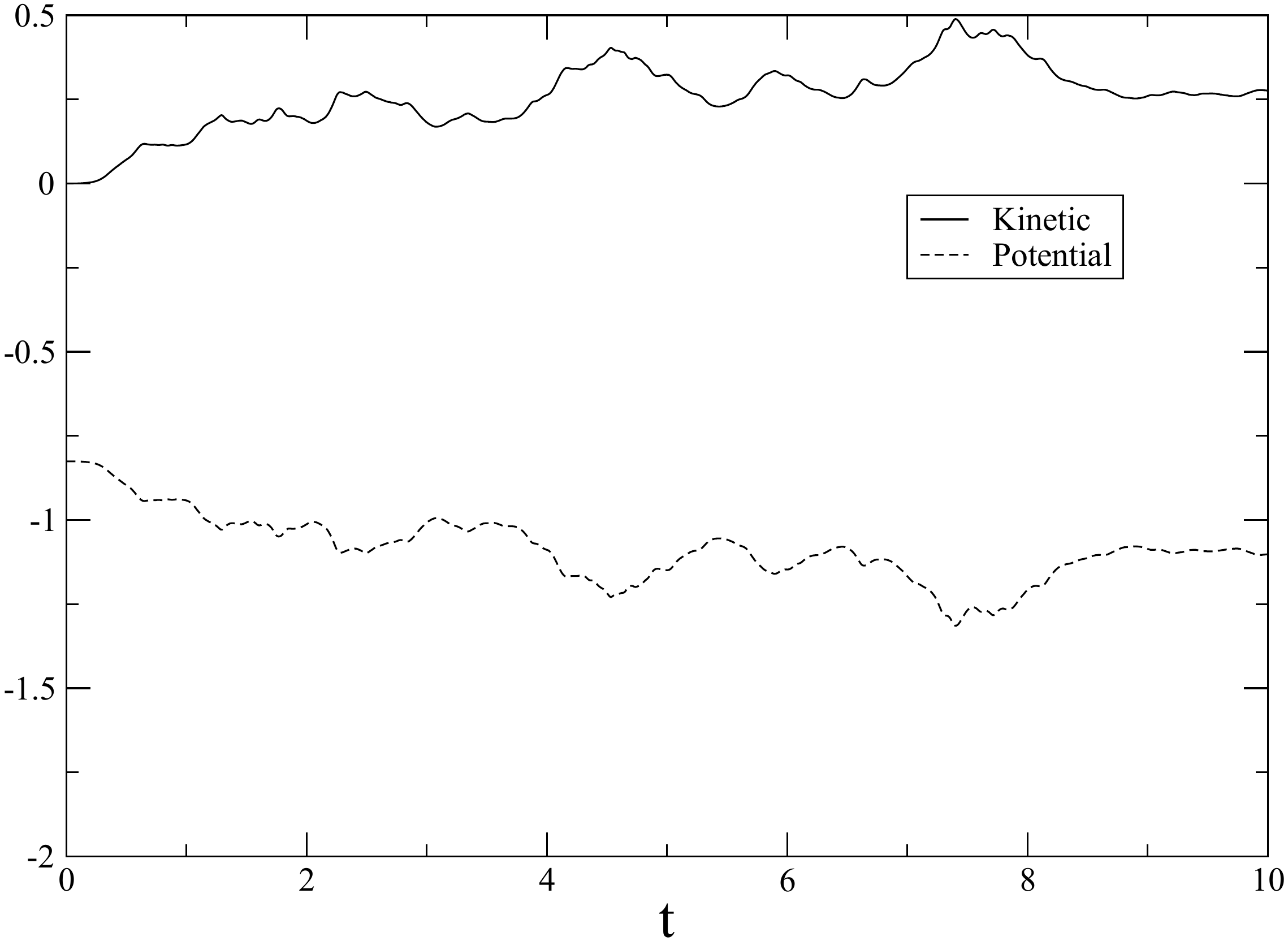}}}
\scalebox{0.25}{{\includegraphics{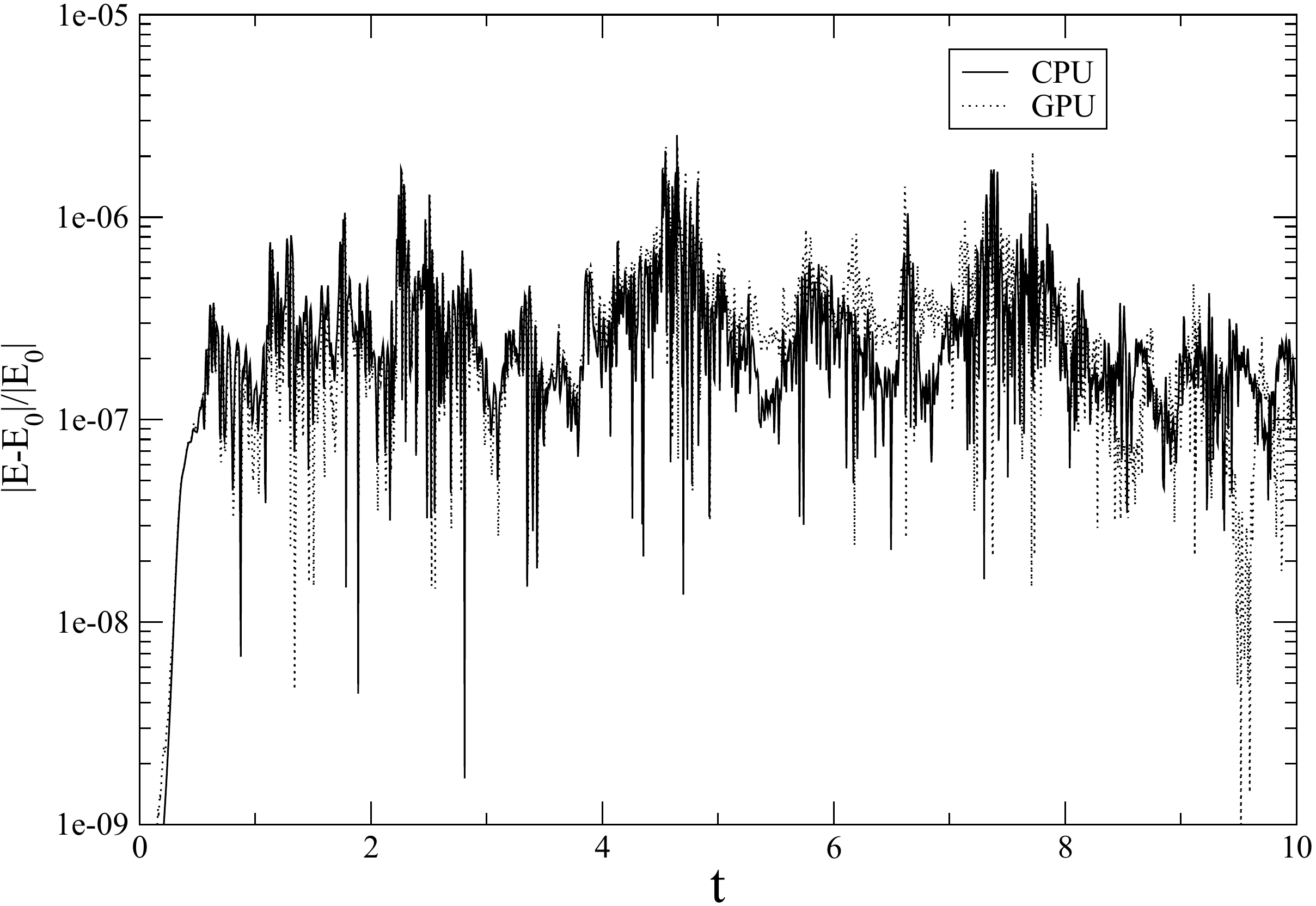}}}
\end{center}
\caption{Left panel: Kinetic and potential energy as a function of time for the Ring model with $\epsilon=10^{-3}$, $\Delta t=0.005$.
Right panel: Error for the CPU implementation in full double precision and the GPU implementation.}
\label{fig5}
\end{figure}

\begin{figure}
\begin{center}
\scalebox{0.25}{{\includegraphics{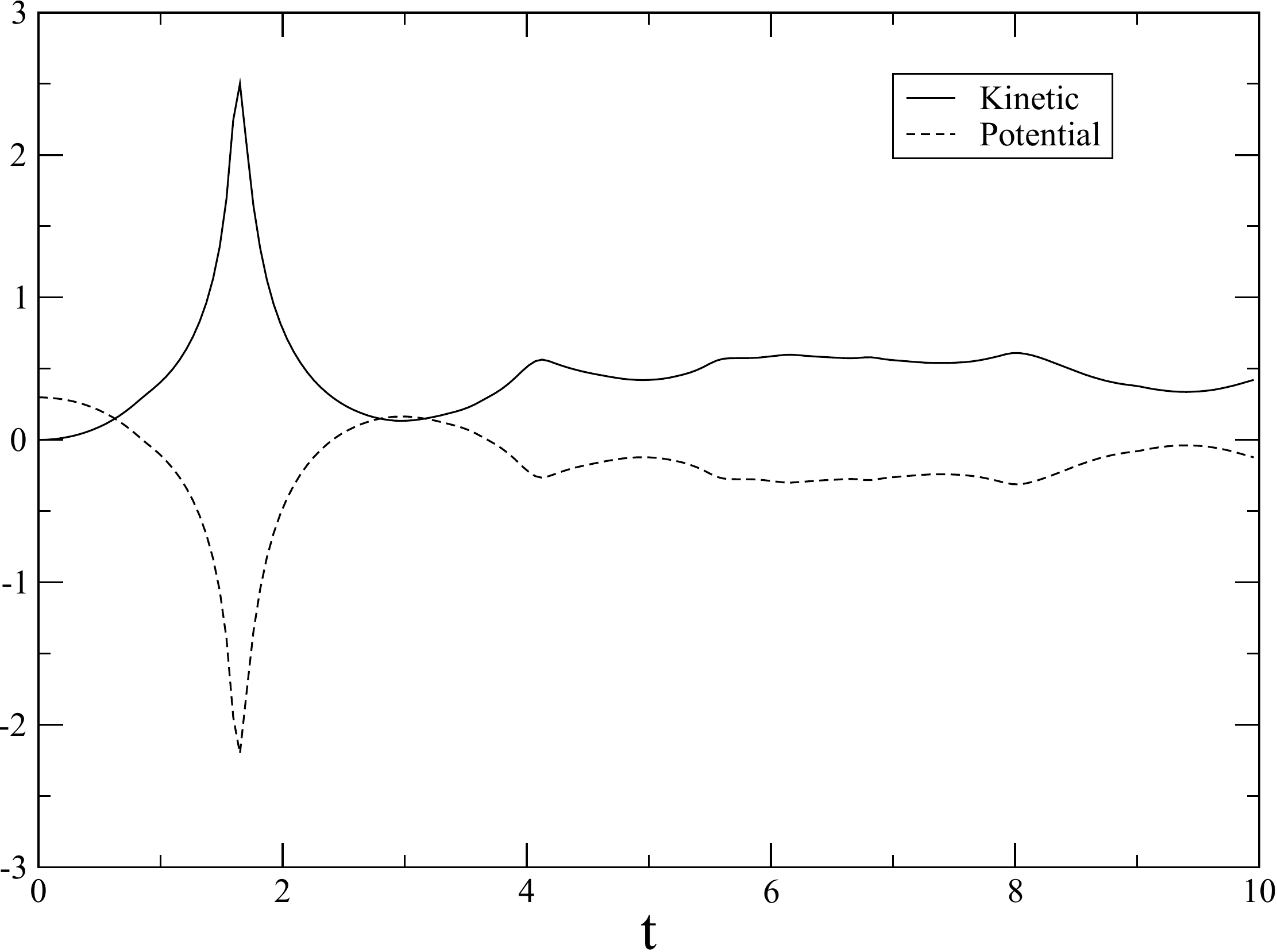}}}
\end{center}
\caption{Kinetic and potential energy as a function of time for the two-dimensional self-gravitating system, with $\epsilon=10^{-3}$, $\Delta t=0.005$.}
\label{fig7}
\end{figure}

\begin{figure}
\begin{center}
\scalebox{0.24}{{\includegraphics{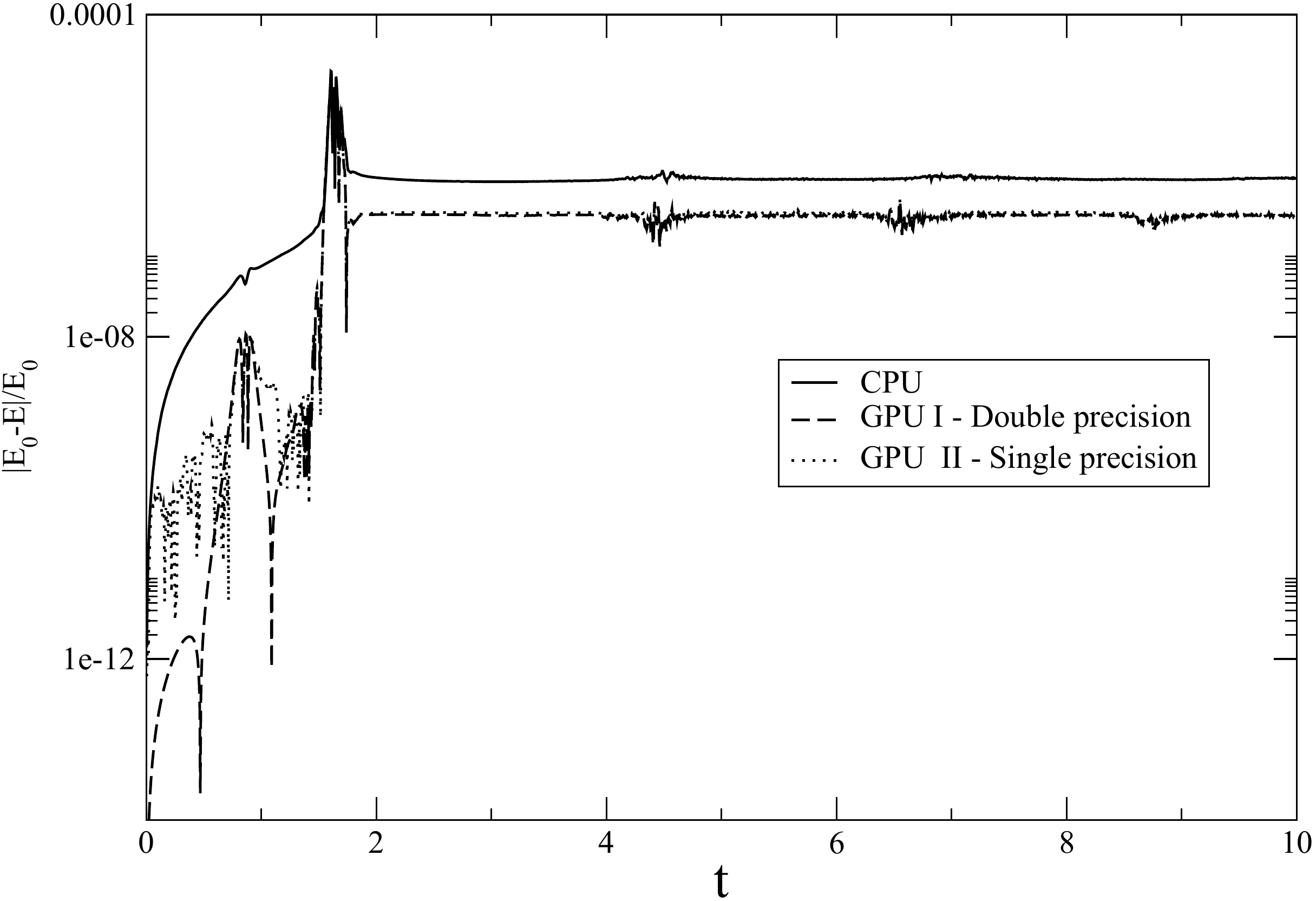}}}
\scalebox{0.24}{{\includegraphics{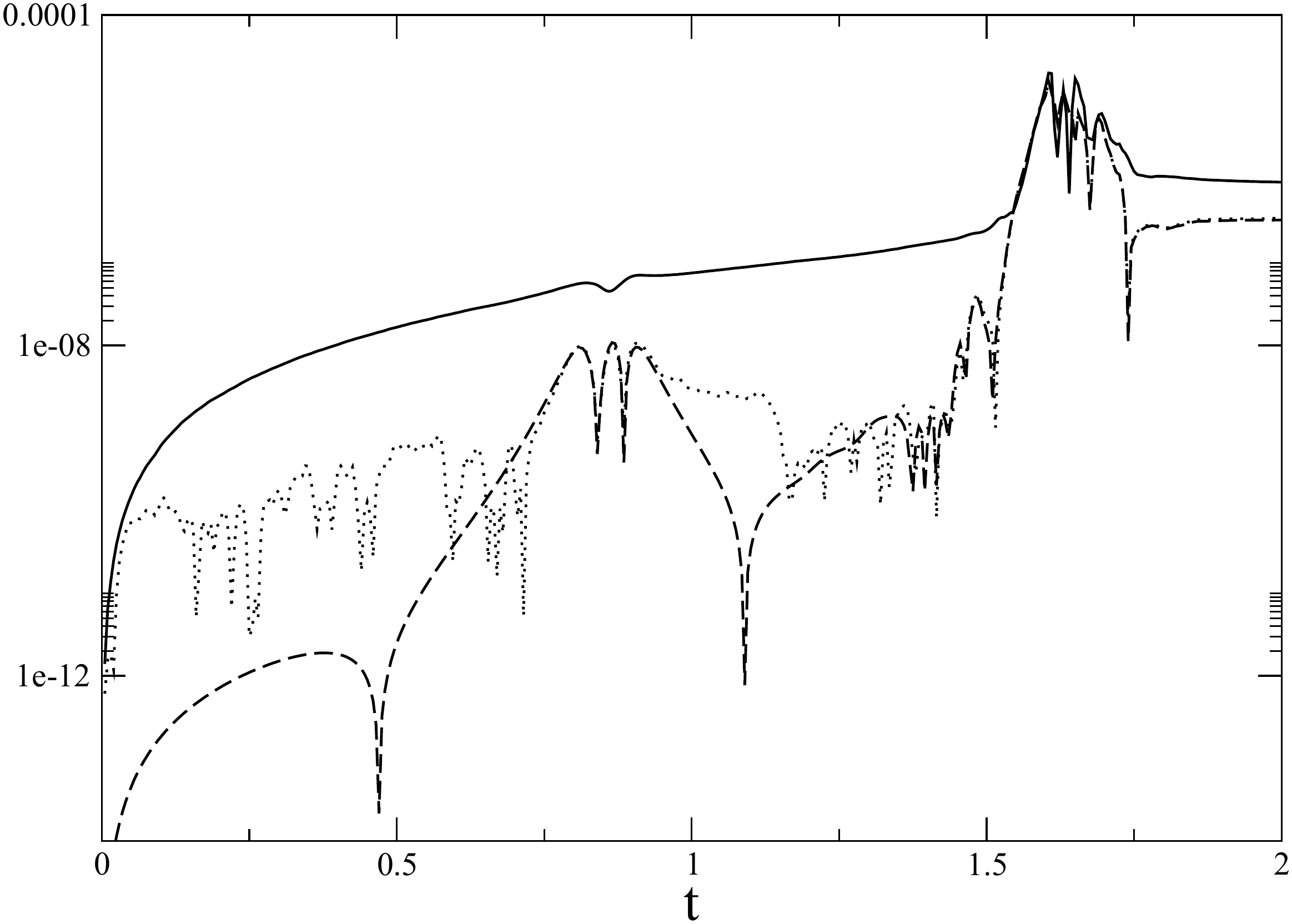}}}
\end{center}
\caption{Relative error in the total energy for the self-gravitating system for the CPU and GPU implementations. For comparison purposes all computations
on the CPU implementation were performed in double precision. The right panel is a zoom over the region comprising the violent relaxation stage.}
\label{fig8}
\end{figure}

To assess the accuracy of the present approach, we consider all three systems and compare the relative error in the energy $\varepsilon=|(E-E_0)/E_0|$ for
the different implementations. For the sake of comparisons, the error for the CPU case is obtained using double precision in all steps.
For all three system we consider an initial state with all particles initially at rest. For the HMF and Ring
models the initial state is spatially homogeneous. Figures~\ref{fig3} and~\ref{fig5} show the plot of the kinetic and potential energies and the
corresponding error $\varepsilon$ for a time window large enough to encompass the initial violent relaxation of the system which is more pronounced for the
HMF model. For the latter, the error for the CPU and GPU implementation are indistinguishable in the plot. For the Ring model  both cases are always very close.
Thence computing the force between the two particles using single precision arithmetic do not compromise the overall accuracy.
For the two-dimensional self-gravitating system, we consider all particles initially at rest and homogeneously
distributed on a circular shell of inner and outer radius $R_1=1.0$ and $R_2=1.5$. The system then undergoes a
violent relaxation towards a quasi-stationary state with some damping oscillations as show in Fig.~\ref{fig7}. Figure~\ref{fig8} shows the plot of the error for
a CPU implementation (in FORTRAN) using double precision for all steps, and the error for the GPU implementation also in full double precision, and
GPU implementation using single precision for the computation of forces between pairs of particles, as described above yielding the same order of
magnitude for the error. The error grows as the particles
get closer during the violent relaxation with a threshold at the same values for the two GPU implementations.

\begin{table}
\begin{center}
\begin{tabular}{c|c|c|c}
\hline\hline
$N$ & CPU & Single & Dual\\
\hline
\hline
 $10^5$ & 0.014 & 6.3$\times 10^{-4}$ & 6.2$\times 10^{-4}$\\
 $10^6$ & 0.14 & 4.0$\times 10^{-3}$ & 2.2$\times 10^{-3}$\\
 $10^7$ & 1.40 & 3.7$\times 10^{-2}$ & 2.0$\times 10^{-2}$ \\
 $5\times10^7$ & 6.90 & 1.81$\times 10^{-1}$ & 9.5$\times 10^{-2}$ \\
\hline
\end{tabular}
\end{center}
\caption{Run times in seconds for a single complete time step for the HMF model with $N$ particles for CPU, and the GPU used as single and dual units.
}
\label{tab1}
\end{table}

\begin{table}
\begin{center}
\begin{tabular}{c|c|c|c|c}
\hline\hline
$N$ & CPU & Single & Dual & Tile size\\
\hline
\hline
 $10,240$ & 2.8 & 0.04 & 0.027 & 256 \\
 $51,200$ & 65.0 & 0.95 & 0.54 & 512 \\
 $102,400$ & 250.4 & 3.9 & 1.98 & 512 \\
 $512,000$ & 6,726.9  & 97.5 & 47.8 & 1024 \\
\hline
\end{tabular}
\end{center}
\caption{Run times in seconds per time step for the Ring model with $N$ particles .}
\label{tab2}
\end{table}

\begin{table}
\begin{center}
\begin{tabular}{c|c|c|c|c}
\hline\hline
$N$ & CPU & Single & Dual & Tile size\\
\hline
\hline
10,240 & 0.58 & 0.023 & 0.011 & 256 \\
20,480 & 2.48 & 0.047 & 0.025 & 256 \\
102,400 & 57.6 & 0.99 & 0.58 & 512 \\
512,000 & 1,491.9 & 26.8 & 17.42 & 512\\
1,024,000 & 6,455.9 & 106.5 & 64.6 & 1024 \\
\hline
\hline

\hline
\end{tabular}
\end{center}
\caption{Run times per time step in seconds for the 2D self-gravitating system.
}
\label{tab3}
\end{table}

\section{Conclusions}
\label{conc}

We presented implementations in one and two GPUs of a forth order symplectic integrator for three different systems with long-range interactions.
Those systems
have been extensively studied in the literature and the present implementations allowed a more extensive investigation of properties
such as the nature of non-equilibrium phase transitions and non-ergodic behavior in the HMF model~\cite{nos3,nos4}. Shared memory is used
to allow faster memory access crucial in GPU implementations. Since it is a very limited resource using single precision floating point
storage optimizes its use. Some steps of the computation, such as the force between pairs of particles, or the inverse of the interparticle distance,
can be performed in single floating point precision without compromising the overall error in the simulation. The speedups obtained
range typically from 30 to 140 depending on the system and the number of particles, simulations that would be unfeasible
in any reasonable time using a typical CPU. The generalization of the present approach to multi-GPU (more than two) and other $N$ particle
systems is straightforward.

\section{Acknowledgments}

The author would like to thank CNPq and CAPES (Brazil) for partial financial support.

\end{document}